\documentclass{ws-procs10x7}
\usepackage{balance}

\def\de{\Delta E}
\def\mbc{m_{\rm bc}}
\def\ebeam{E_{\rm beam}}

\def\ipb{{\rm pb}^{-1}}

\def\mmiss{\rm M_{miss}}
\def\bf{{\cal{B}}}

\newcolumntype{d}[1]{D{.}{.}{#1}}

\makeindex
\begin{document}

\title{Cabibbo-Allowed and Doubly Cabibbo Suppressed $D\to K\pi$ Decays}

\author{Steven R. Blusk}

\address{Department of Physics, Syracuse University, Syracuse, NY 13244\\
E-mail: sblusk@phy.syr.edu}


\twocolumn[\maketitle\abstract{We present measurements of the branching
fractions of the decays, $D\to K^0_{S,L}\pi$. The measured asymmetry 
shows that $\bf(D^0\to K^0_S\pi^0)\ne\bf(D^0\to K^0_L\pi^0)$, as
expected. We also find that $\bf(D^+\to K^0_S\pi^+)$ is statistically compatible with $\bf(D^+\to K^0_L\pi^+)$.
Lastly, we present a recent measurement of the branching fraction of the doubly Cabibbo-suppressed decay, $D^+\to K^+\pi^0$.}
]

\section{Introduction}

For over two decades, the $D^0\to K^-\pi^+$ has served as a workhorse
in charm and beauty physics. However, there is general interest
in measuring all the $D\to K\pi$ branching fractions. In particular,
while it is often assumed that ${\cal{B}}(D\to K^0_S\pi)={\cal{B}}(D\to K^0_L\pi)$,
interference between $D\to K^0\pi$ and $D\to \overline{K^0}\pi$ can break this
equality\cite{bigi}. Although this asymmetry is expected, measuring it has alluded
experiments because of the challenge of reconstructing the $K^0_L$. Another
$K\pi$ mode which has alluded experiments is the doubly Cabibbo-suppressed (DCS)
$D^+\to K^+\pi^0$. The difficult stems primarily from the low rate, but
also for hadron machines, the lack of a detectable displaced vertex and the large
$\pi^0$ combinatorial background make this mode extremely difficult to detect.
These difficult $D\to K\pi$ modes are accessible at CLEO-c due to the
low-multiplicity environment and threshold production of $D\bar{D}$.

    The analyses presented are based on a 281~$\ipb$ sample of data collected
at the peak of the $\psi(3770)$ ($\sqrt{s}=3774$ MeV). The resonance is just
above threshold for production of $D\bar{D}$, and therefore the final state
is in a coherent C=-1 state. For $D^0\bar{D^0}$, these quantum correlations 
produce deviations in measured branching fractions\cite{asner_sun}, which 
are maximal when CP eigenstates, $S_{\rm CP}$, are involved. For example, the rate 
for ($D^0\to S_{\pm},\bar{D}^0\to S_{\pm}$) is zero, and 
($D^0\to S_{\pm},\bar{D^0}\to S_{\mp}$) is twice as large with respect
to the values obtained when quantum correlations are absent. 
Four cases of interest that enter into the analyses presented here are:
($D^0\to S_{CP\pm},\bar{D^0}\to X$) and ($D^0\to S_{\pm},\bar{D^0}\to f$), where
$f$ represents a flavored final state and $X$ is an unspecified final state. 
Because of the quantum correlations, the branching fractions are modified as 
shown in Table~\ref{tab:cpcorr}\cite{asner_sun}, where $x$ and $y$
are the mixing parameters, $r_f e^{-i{\delta_f}}\equiv<f|\bar{D^0}>/<f|D^0>$,
and $z_f\equiv\cos\delta_f$. In untagged analyses we can easily correct branching
fractions using the word-average $y=0.008\pm0.005$\cite{pdg04}. We also note that yields in 
these and other combinations of final states can be used to measure
the $D^0\bar{D}^0$ mixing parameters and the strong phase $\delta_{K\pi}$\cite{asner_sun}.

\begin{table}
\tbl{Quantum correlation factors for four $D^0\bar{D^0}$ final state
configurations.\label{tab:cpcorr}}
{\begin{tabular}{@{}lcc@{}} 
\toprule
               & $S_+$                            &      $S_-$       \\
\hline
$f$            & $1+2r_fz_f+r_f^2$      & $1-2r_fz_f+r_f^2$ \\
$X$            & $1-y$                  & $1+y$  \\
\botrule
\end{tabular}}
\end{table}

    In reconstructing $D$ mesons, we use two kinematic variables:
$\de\equiv E_{D}-\ebeam$ and $\mbc\equiv\sqrt{\ebeam^2-p_D^2}$, where $E_D$
is the energy of the $D$ candidate and $p_D$ its momentum. {\it Untagged}
analyses reconstruct $D$ mesons in exclusive final states using all 
charged particles and showers in the event. {\it Tagged} analyses start
with events that already have a $D$ candidate {\it ie., a tag}, 
and seek to reconstruct the second $D$ meson (referred to as the {\it signal}). 
Because of the highly constrained kinematics, the signal $D$ may contain
undetected particles, such as a $K^0_L$ (or a $\nu$), which are inferred
by energy/momentum conservation. In particular, for the decay $D\to K^0_L\pi$, the
signal is a peak in the missing-mass squared, defined using the measured
four-momenta as: $\mmiss^2 = (p_{\rm event}-p_{tag}-p_{\pi})^2$.

\section{$\bf(D^0\to K^0_{S,L}\pi^0)$} 

     We first measure $\bf(D^0\to K^0_S\pi^0)$ using an untagged analysis.
Candidates are formed by combining $K^0_S\to\pi^+\pi^-$
and $\pi^0$ candidates and requiring $\de$ and $\mbc$ to be within
3 standard deviations of 0 and $M_{D^0}$, respectively. Combinatorial
background and cross-feed from $D^0\to\pi^+\pi^-\pi^0$ are estimated
using $\de$ and $K^0_S\to\pi^+\pi^-$ mass sideband regions, respectively. 
Combining the signal yield of $7487\pm99$ events with the
efficiency of 29.0\% and $N_{D^0\bar{D^0}}=1.015\times10^6$, we find:
$\bf(D^0\to K^0_S\pi^0)=(1.260\pm0.02\pm0.054)\%$. Of the 4.2\% systematic
uncertainty, 3.8\% is from the $\pi^0$ detection efficiency, which 
cancels when comparing $K_S^0\pi^0$ and $K_L^0\pi^0$.

    Measurement of $\bf(D^0\to K^0_L\pi^0)$ requires a tagged analysis, and
since $K^0_L\pi^0$ is a CP+ eigenstate, it requires that we determine
the factor $1+2r_fz_f+r_f^2$ (which is unknown, since $\delta_f$ is unknown). 
However, by measuring $\bf(D^0\to K^0_S\pi^0)$ in tagged events, and comparing
to the value in untagged events, we can determine $(1-2r_fz_f+r_f^2)$. 
Along with the measured values of $r_f$, 
this enables us to compute the factor we want, $(1+2r_fz_f+r_f^2)$. We therefore
need $\bf(D^0\to K^0_S\pi^0)$ in flavor-tagged events.

The tagged $D^0\to K^0_S\pi^0$ tagged analysis starts with events
containing a reconstructed $D$-tag in $\bar{D^0}\to K^+\pi^-$, 
$\bar{D^0}\to K^+\pi^-\pi^0$ or $\bar{D^0}\to K^+\pi^-\pi^+\pi^-$, and 
then seeks to reconstruct $D^0\to K^0_S\pi^0$ candidates as described
in the untagged analysis. The yields, efficiencies and corresponding
products $\bf(D^0\to K_S^0\pi^0)(1-2r_fz_f+r_f^2)$ are shown in
Table~\ref{tab:kspi0_sum}. Using the measured value of $\bf(D^0\to K^0_S\pi^0)$
from the untagged analysis, we also compute $(1-2r_fz_f+r_f^2)$ and
subsequently $(1+2r_fz_f+r_f^2)$ using the most recent 
$r_f$ values\cite{pdg04,zhang,tian}. That these factors are not unity 
is a direct consequence of the quantum coherence of the final state.

\begin{figure}
  \includegraphics[height=.23\textheight]{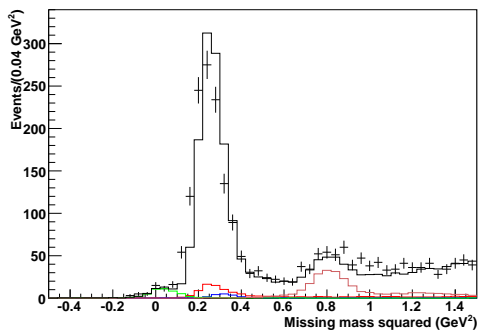}
  \caption{Distribution of $\mmiss^2$ for $D^0\to K^0_L\pi^0$ candidates in tagged
events. The points with error bars are data, the solid line is the total simulation,
and the dashed lines are various backgrounds.
\label{fig:klpi0_mmiss}}
\end{figure}

The measurement of $K^0_L\pi^0$ is slightly more complicated. It starts
with the same sample of D-tag's as in the $K^0_S\pi^0$ tagged analysis, 
and, for each candidate, we require the presence of one and 
only one additional $\pi^0$ candidate, and no extra tracks or
$\eta\to\gamma\gamma$ candidates. In these events, we form 
$\mmiss^2$, which for $K^0_L\pi^0$ events peaks at $M_{K_L^0}^2$.
Backgrounds such as $K_S^0\pi^0$ and $\eta\pi^0$ are highly suppressed
by the selection requirements, but do peak under the signal. These
backgrounds are estimated using simulation. Other backgrounds
are estimated using $\mmiss^2$ sidebands in data in combination with 
shapes from simulation. The distribution of $\mmiss^2$ is shown in
Fig.~\ref{fig:klpi0_mmiss} for all tag modes combined; 
the data are the points with error bars,
the solid line is the simulation, and the dashed lines show various 
background contributions. The data are peaked toward slightly lower missing-mass
than simulation. This effect is traced to a 0.5\% difference in the energy 
scale of $\pi^0$'s, which has only a minor effect in this analysis.
Yields, efficiencies and the branching fractions,
$\bf(D^0\to K_L^0\pi^0)$, are shown in Table~\ref{tab:klpi0_sum}, where
the branching fractions have been corrected by the factor, $(1+2r_fz_f+r_f^2)$.
After averaging the three tag modes, we obtain 
$\bf(D^0\to K_L^0\pi^0)=(0.986\pm0.049\pm0.047)\%$, where the last uncertainty
is systematic and dominated by the $\pi^0$ efficiency (3.8\%).

Defining an asymmetry:
\begin{equation*}
R(D) = \frac{\bf(D\to K_S^0\pi)-\bf(D\to K_L^0\pi)}{\bf(D\to K_S^0\pi)+\bf(D\to K_L^0\pi)},
\end{equation*}

\noindent we find that $R(D^0)=0.122\pm0.025\pm0.019$, establishing the inequality of these
branching fractions. Using general arguments involving the contributing 
Feynman diagrams, one would expect this asymmetry to be 
$R(D^0)=2\tan^2\theta_C=0.109\pm0.001$, where $\theta_C$ is the Cabibbo angle.
This expectation is in good agreement with our measurement.

\begin{table*}
\tbl{Summary of results for the $D^0\to K^0_S\pi^0$ tagged analysis. 
\label{tab:kspi0_sum}}
{\begin{tabular}{@{}lccc@{}} 
\toprule
Tag Mode ($f$)            &  $K^+\pi^-$     &  $K^+\pi^-\pi^0$  & $K^+\pi^-\pi^+\pi^-$   \\
\hline
Tag Yield                 &  47440          &   64280           &  75113                 \\
Signal Yield              &  155            &    203            &  256                \\
Efficiency (\%)            &  31.47          &    31.45          &  30.69                 \\
\hline
$\bf(D^0\to K_S^0\pi^0)\times$  &                 &                   &                        \\
$(1-2r_fz_f+r_f^2)$ (\%) & $1.04\pm0.09$   & $1.01\pm0.09$     &  $1.17\pm0.08$ \\
\hline
$(1-2r_fz_f+r_f^2)$      &$0.824\pm0.013\pm0.073$ & $0.802\pm0.013\pm0.068$ & $0.932\pm0.015\pm0.063$ \\
\hline
$(1+2r_fz_f+r_f^2)$      & $1.183\pm0.013\pm0.073$ & $1.203\pm0.013\pm0.068$ & $1.074\pm0.015\pm0.063$ \\
\botrule
\end{tabular}}
\end{table*}

\begin{table*}
\tbl{Summary of results for the $D^0\to K^0_L\pi^0$ tagged analysis. 
\label{tab:klpi0_sum}}
{\begin{tabular}{@{}lccc@{}} 
\toprule
Tag Mode ($f$)            &  $K^+\pi^-$     &  $K^+\pi^-\pi^0$  & $K^+\pi^-\pi^+\pi^-$   \\
\hline
Tag Yield                 &  47440          &   64280           &  75113                 \\
Signal Yield              &  334.8          &    363.1          &  418.0                \\
Efficiency (\%)            & 55.21           &    54.67          &  52.72                 \\
\hline
$\bf(D^0\to K_L^0\pi^0)$ (\%)  & $1.029\pm0.011\pm0.088$ & $0.818\pm0.009\pm0.067$  & $0.990\pm0.014\pm0.079$ \\
\botrule
\end{tabular}}
\end{table*}

\section{$\bf(D^+\to K^0_{S,L}\pi^+)$} 

\begin{figure}
  \includegraphics[height=.23\textheight]{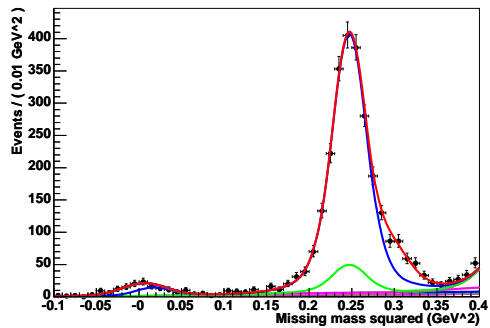}
  \caption{Distribution of $\mmiss^2$ for $D^+\to K^0_L\pi^+$ candidates in tagged
events. The points with error bars are data, the solid line are signal and background
contributions as described in the text.
\label{fig:klpi_mmiss}}
\end{figure}

We look to measure the same asymmetry in charged $D$ decays. The branching
fraction, $\bf(D^+\to K^0_S\pi^+)$ has been measured in a separate analysis\cite{dhad}.
The measurement of $\bf(D^+\to K^0_L\pi^+)$ requires a tagged analysis, and
is strategically similar to the $\bf(D^0\to K^0_L\pi^0)$ measurement. We reconstruct
a $D^-$ tag in 6 tag modes: $D^-\to K^+\pi^-\pi^-$, $K^+\pi^-\pi^-\pi^0$, $K^0_S\pi^-$,
$K^0_S\pi^-\pi^0$, $K^0_S\pi^-\pi^+\pi^-$, and $K^+K^-\pi^-$, by requiring $\de$
consistent with zero. Selecting events within $\sim$3$\sigma$ of $M_{D^-}$,
we obtain 165,00 $D^-$ tags. For each tag,
we query the remainder of the event and require exactly 1 extra charged 
track, consistent with a pion hypothesis, and no extra $\pi^0$'s. Using the $D^-$ 
tag and the pion, we compute $\mmiss^2$, which is shown in Fig.~\ref{fig:klpi_mmiss}
for all tag modes combined.
The points with error bars show the data, and the colored line passing through the
points shows the fit. The prominent $K^0_L$ peak is evident as well as a high-side 
shoulder from $D^+\to\eta\pi^+$ (this analysis does not veto $\eta\to\gamma\gamma$).
The other lines show the individual contributions from $K^0_L\pi^+$ (signal),
and various backgrounds, such as $K^0_S\pi^+$, which peaks under $K^0_L\pi^+$; 
$D^+\to\pi^+\pi^0$ and $D^+\to\mu^+\nu_{\mu}$, which peak near zero; and other non-peaking
backgrounds. A total of 2023$\pm$54 $D^+\to K^0_L\pi^+$ signal events are observed
from an initial tagged sample of 165,000 $D^-$ tags.

The branching fraction is computed for each tag mode and then the results are
combined using a weighted average. The efficiency varies slightly depending on 
the tag mode, but is typically about 82\%. The average branching fraction is
found to be: $\bf(D^+\to K^0_L\pi^+)=1.46\pm0.040\pm0.035\pm0.004)$, where the
last systematic is due the uncertainty in $\bf(D^+\to K^0_S\pi^+)$.

\newpage
Using $\bf(D^+\to K^0_S\pi^+)=(1.552\pm0.022\pm0.029)\%$\cite{dhad}, we measure
an asymmetry, $R(D^+)=0.031\pm0.016\pm0.016$. This asymmetry is consistent with 
zero. Because of the larger number of additional Feynman diagrams which contribute
to this decay, no simple prediction of this asymmetry can be made. 
Both this analysis and the $D^0\to K^0_{S,L}\pi^0$ will be submitted for publication
soon.

\section{$\bf(D^+\to K^+\pi^0)$}

    Until recently, the DCS $D^+$ decays were limited to modes with only charged
particles due to the low rate and large combinatorial background associated with 
$\pi^0$ reconstruction. The threshold production of $D\bar{D}$ events in CLEO-c
make this measurement accessible\cite{cleo_dcs}. CLEO searches for this decay using an untagged 
analysis by combining $K^+$ and $\pi^0$ candidates and requiring $-40<\de<35$~MeV.
We find a yield of $148\pm23$ events. We use $D^+\to K^-\pi^+\pi^+$ as a normalizing
mode, for which there are 79612 decays. The efficiencies of the DCS and normalizing
mode are 42.30\% and 52.16\%, respectively, yielding a branching fraction,
$\bf(D^+\to K^+\pi^0)=(2.28\pm0.36\pm0.15\pm0.08)\times10^{-4}$. This measurement is
of substantially better precision than a recent measurement by BaBar\cite{babar}, 
which used data collected near the $\Upsilon$(4S) with $\sim$1000 times larger 
integrated luminosity than CLEO-c.

We gratefully acknowledge the effort of the CESR staff 
in providing us with excellent luminosity and running conditions,
and the National Science Foundation for support of this work.

\end{document}